 \documentstyle[aps,prb,epsf,multicol]{revtex}  

\newcommand\dg{^{\dag}}

\def\beq{\begin{equation}}
\def\eeq{\end{equation}}
\def\bea{\begin{eqnarray}}
\def\eea{\end{eqnarray}}

\begin{document}
\draft

\title{Two-channel Kondo Lattice Model on a ladder studied by the 
Density Matrix Renormalization Group Method}
\author{Juana Moreno$^{1}$, Shaojin Qin$^{2}$, P. Coleman$^{3}$,
 and Lu Yu$^{4,2}$}
\address{$^{1}$ Dept of Physics \& Astronomy, Northwestern Univ.,
2145 Sheridan Rd, Evanston IL 60208, USA }  
\address{$^{2}$  Institute of Theoretical Physics, P. O. Box 2735, 
Beijing 100080, People's Republic of China}
\address{$^{3}$  Serin Laboratory, Rutgers University,  
136 Frelinghausen Road, Piscataway, NJ 08854-8019, USA.} 
\address{$^{4}$ International Center for Theoretical Physics, 
P.O. Box 586, 34100 Trieste, Italy}

\date{\today}
\maketitle
\begin{abstract}
Using the density matrix renormalization group (DMRG) method 
we study a two-channel Kondo lattice model on a half filled 
ladder. Our model involves an on-site s-wave and a nearest neighbor 
d-wave coupling between the local moments and the conduction electrons 
on the ladder. By changing the relative strength of the two 
Kondo interactions we examine the evolution of the system from 
a conventional  Kondo insulator with a singlet at each site to a new 
kind of semimetallic state formed by overlapping of 
Zhang-Rice-like singlets.
The DMRG is used to  study how the spin and charge correlation functions 
evolve between these two regimes.
\end{abstract}
\begin{multicols}{2}

\section{\bf Introduction}

The interaction between localized spins and delocalized conduction electrons 
plays a central role in the physics of heavy fermion 
materials,\cite{hfermion} Kondo insulators \cite{Riseborough,fisk} and many 
other highly correlated electron systems.  The Kondo lattice 
model (KLM) is probably the simplest theoretical model that describes
this physics. 

Considerable insight into local moment materials has been obtained
by studying the one-dimensional Kondo lattice. 
\cite{Ueda}
Although some properties are intrinsic to 
one dimensional models, such as the nesting instabilities 
\cite{organic-review}, other aspects such as the insulating behavior
at half filling and the Fermi surface expansion by the local moments
\cite{Oshikawa} are also inherent to three-dimensional systems.
Recently, the theoretical study of one-dimensional models has also proved
a most useful tool in the analysis of an increasing 
number of low-dimensional materials.
The comparison of experimental findings with  exact numerical 
results  helps to evaluate the adequacy of  basic models 
and to quantify the relevance of additional effects 
such as the importance of the orbital electronic
degrees of freedom. 
\cite{Yamada,Cox-review}

The earliest claims of quasi-one-dimensional behavior in Kondo 
lattice intermetallic compounds date back to the seventies, 
\cite{Zn13X,Al11Mn4} but only recently,
with the synthesis of the quasi-one-dimensional 
molecular conductor 
$(DMET)_2FeBr_4$, has  a clear-cut experimental realization of a 1D Kondo 
lattice been found. \cite{DMET2FeBr4} This compound displays 
features of a strongly correlated electron system. It shows a 
metal-insulator transition at 200 K, which appears to be driven by
the interaction between the $\pi$-conduction electrons of the 
$DMET$ donor sheets and the localized $Fe^{3+}$ spins. \cite{char-DMET}

A class of heavy fermion systems, known as Kondo insulators, 
display insulating properties at low temperatures.
\cite{Riseborough,fisk} One of the interesting aspects of a Kondo lattice is the emergence
of insulating ground-states. 
The ground state of the conventional 
Kondo lattice model is a non-magnetic insulator at half-filling,
with a gap in both spin and charge excitation spectrum.\cite{Ueda}
In this way, the half-filled Kondo lattice provides 
a simple model for the Kondo insulator, where the coherent Kondo 
effect forms a highly renormalized band-insulator.
However, the smallest gap Kondo insulators, $CeNiSn$ and $CeRhSb$, 
do not fit into this simple scheme: they appear to develop gapless 
excitations. \cite{CeNiSn-conduc} The narrow-gap Kondo insulators 
can be regarded as an example of a half-filled Kondo system with a
gapless  ground state due to the presence of interactions that are 
absent in the conventional Kondo lattice model. \cite{CeNiSn,Miyake}

A simple model which displays a metallic ground state at 
half-filling is the ``d-wave Kondo model''. This is a variant of 
the conventional Kondo model where each local spin couples through 
the exchange interaction  with a d-wave orbital formed by 
the nearest-neighbor electrons.  This interaction favors the formation 
of Zhang-Rice-like singlets \cite{Zhang-Rice} between each 
spin and the nearest conduction electron orbitals. The superposition
of these extended singlets may establish a new kind of semimetallic 
state, which may be relevant to explaining the metallic phase 
of $(DMET)_2FeBr_4$.  Another interesting feature of the d-wave Kondo 
model is the introduction of spin-dependent hopping processes, where
an electron is able to jump to the next-nearest-neighbor site by flipping 
the spin of the local moment. This kind of interaction plays a role in 
other organic compounds, such as  the magnetic graphite intercalation 
compounds.  \cite{spin-hop}

Motivated by the need to understand how a metal-insulator transition arises 
from pure exchange interactions between the itinerant and localized 
electrons, this paper examines a Kondo model with two spin-exchange channels 
sharing a single conduction band: an on-site s-wave 
(Fig. \ref{fig:s-wave-ladder}) and a nearest neighbor d-wave coupling 
(Fig. \ref{fig:d-wave-ladder}). 
Electron-electron interactions, such as Hund's 
interactions, can induce the opening of new spin-exchange channels 
between the local moments and the conduction 
electrons.\cite{Schork,Piers-Tsvelik,Cox93} 
For an impurity magnetic ion, the Kondo effect develops exclusively 
in the strongest screening channel due to the local 
symmetry which preserves the channel quantum number of scattered electrons. 
In a lattice, however, a conduction electron can change symmetry channels 
as it moves from one spin site to another. Therefore,
the development of a 
Kondo effect in one channel no longer excludes the possibility of a 
Kondo effect developing coherently in other channels.

Our goal is to determine  whether there is a range of exchange 
coupling for which such a two-channel Kondo model displays gapless 
excitations at half-filling.  If this is the case, 
the competition between the d- and s-wave Kondo couplings might 
induce new intermediate phases between the d-wave Kondo ground state 
(Zhang-Rice singlets at each site) and the conventional Kondo ground 
state (s-wave Kondo singlets).  It has been argued that even a 
composite superconducting phase could appear in the KLM if the local 
moments interact with a single conduction band via two orthogonal 
scattering channels with the same spatial parity.\cite{Piers}

In highly correlated electron systems, it is essential to go 
beyond the mean-field level to understand their low-energy 
properties. We apply the  density matrix renormalization-group 
(DMRG) method \cite{White} to calculate the energy spectrum and 
ground state correlations of this two-channel KLM.  This real-space 
technique has proved to be remarkably accurate for one-dimensional 
systems such as the Kondo and Anderson lattices.
\cite{Yu93,Shibata96,Guerrero} To check the character and energy 
range of the elementary excitations we have also performed a variational 
calculation in the d-wave limit. 

Section II introduces the  Hamiltonian and the parameters of 
our model, the scaling law used to analyze the results and the 
numerical technicalities of the calculation.
We present our results for the ground state energy and the spin and 
charge excitation gaps in section III. In Section IV, we discuss the 
spin and charge correlations of the ground state and some 
characteristics of the excited states. Section V is dedicated to 
a comparison between our results and a variational calculation 
and finally, section VI presents our conclusions.

\section{\bf Model}

\begin{figure}[h]
 \begin{minipage}{\linewidth}
\centerline{\epsfxsize8cm\epsffile{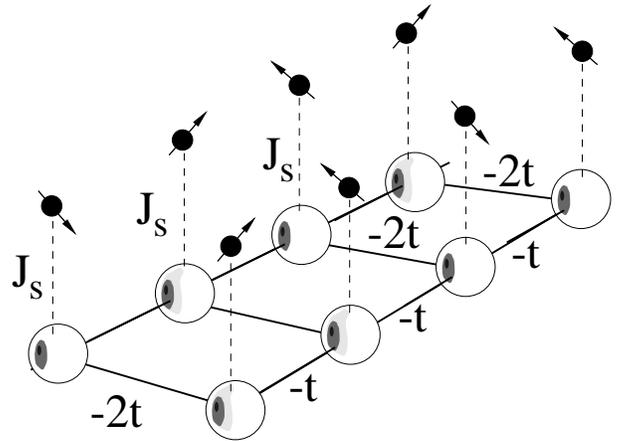}}
\vskip 0.4truein
\caption[]{Ladder illustrating the electronic hopping 
and the s-wave Kondo interaction between each local spin 
(arrows) and the conduction electron at the same site (spheres).}
\label{fig:s-wave-ladder}
 \end{minipage}
\end{figure}

\begin{figure}[h]
 \begin{minipage}{\linewidth}
\centerline{\epsfxsize8cm\epsffile{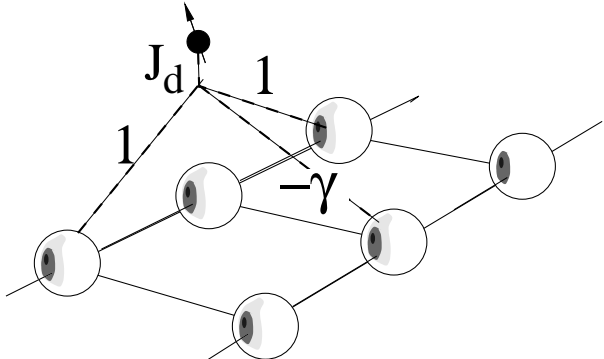}}
\caption[]{Ladder illustrating the d-wave Kondo interaction between 
each local spin and the three near-neighboring electrons. Note that 
the coupling is $J_d$ along the chains and $-\gamma J_d$ along the 
rungs of the ladder.}
\label{fig:d-wave-ladder}
 \end{minipage}
\end{figure}

The minimal one dimensional model with two Kondo exchange channels of 
different symmetry and the same parity is a ladder model.
The exchange interactions between the local moments and the electrons
are an on-site s-wave (Fig. \ref{fig:s-wave-ladder}) and a nearest 
neighbor d-wave coupling (Fig. \ref{fig:d-wave-ladder}). Unlike other 
multichannel Kondo models,\cite{Cox-review} our model involves a 
single conduction electron band and Kondo singlets can form 
for any value of the couplings.
Electrons and local moments are located at sites
$\vec R_i^k = i  \hat x + k \hat y, \hspace{0.1in}
\rm{where}\hspace{0.1in} k=0,1$. 
We impose open boundary conditions in the $x$ direction.  Along the 
$y$ axis, the system can be considered a periodic lattice with a 
period of two lattice spacings. The Hamiltonian is:

\beq
H= H_0+ H_s +H_d,
\eeq
where $H_0$ is the electronic kinetic energy given by,
\bea
H_0 = -t \sum_{i,k} ( \psi^{k\dagger}_{i+1 \alpha} \psi^k_{i \alpha}
                  +\psi^{k+1\dg}_{i \alpha} \psi^{k}_{i\alpha} 
		  + {\rm h. c}),  
\eea
where $t$ denotes the hopping matrix element, and the operator 
$\psi^{k\dg}_{i \alpha}$ creates a conduction electron on site 
$\vec R_i^k$ with spin $\alpha$. $H_s$ is the conventional s-wave 
Kondo interaction:
\bea
H_s=J_s\sum_{i,k} \Big\{
	\vec S^k_i\cdot \psi^{k\dg}_{i\alpha} 
\frac{\vec \sigma_{\alpha \beta}} {2}
\psi^k_{i \beta} \Big \} \quad\rm{(s-channel)}, 
\eea
where $J_s$ is the strength of the antiferromagnetic s-wave coupling, 
$\vec S^k_i$ and $\vec s^k_i=\psi^{k\dg}_{i\alpha} 
(\vec \sigma_{\alpha \beta}/2)\psi^k_{i \beta}$ 
denote the spin of the local moment and the conduction 
electron at position $\vec R_i^k$, respectively.

The last term, $H_d$, represents the d-wave Kondo channel:
\bea
H_d=J_d\sum_{i,k} \Big\{\vec S^k_i\cdot d^{k\dagger}_{i\alpha}
	\frac{\vec \sigma_{\alpha \beta}} {2} d^{k}_{i\beta}
\Big\}\quad\rm{(d-channel)},\\
\quad\rm{where} 
\indent d^{k\dagger}_{i \alpha}=\frac{(\psi^{k\dg}_{i+1 \alpha}+
\psi^{k\dg}_{i-1 \alpha}-\gamma \psi^{(k+1)\dg}_{i \alpha})}
{\sqrt{2+\gamma^2}}.
\eea
The operator $d^{k\dagger}_{i \alpha}$
creates an electron with ``d-wave 
symmetry'' in the ``$\gamma$'' Wannier state at site $\vec R_i^k$. 
For $\gamma=1$ this state has pure d-wave symmetry
but since the one-dimensional ladder breaks the $90^{0}$
rotation symmetry of the lattice, we explore a more general class of
states  by allowing $\gamma $ to become
a free parameter in the model (in this way our ``d-channel'' involves
an admixture of d and extended s-wave character). Note that
the d-wave spin exchange involves spin-dependent hopping
processes where the electron hops between next-near-neighbor sites via
the exchange of spin with the local moments. 

Our interest lies in the half-filled case where the total number 
of conduction electrons is equal to the number of lattice sites.
Finite size studies on our model are facilitated by using the 
strong-coupling limit to reduce the feasibility of antiferromagnetic 
instabilities, although, as we discuss later, we can not completely 
eliminate them.  Thus, we use the following values of the parameters,
$t=0.01$ and $|J|=\sqrt{J_s^2+J_d^2}=1$. 
A finite hopping $t$ is required to ensure the smooth distribution of the
electronic density. 
We have studied three values of the transverse exchange parameter,
$\gamma=2,1,0.5$, and several values of the ratio between the
d- and s-wave exchange couplings, $J_d/J_s$. 

We use the finite-chain DMRG algorithm \cite{White} up to ladders 
of 12 rungs, using up to  100 optimized states for each block. 
In contrast to the infinite system method, the finite system algorithm 
gives more accurate results, but an asymptotic extrapolation to the 
infinite system is necessary.  We use the following scaling 
law, \cite{Sorensen} which is appropriate for a massive mode,
\begin{equation}
\Delta (L)= 
\sqrt{\Delta^2 (L=\infty) + v^2 \frac{\pi^2}{(L+1)^2}}
\label{scaling}
\end{equation}
where $\Delta(L)$ is the gap for a system of length $L$.  This  
allows us to define a correlation length, 
$\xi=v/\Delta(L=\infty)$, for the spin and charge modes of the system.
In the DMRG calculations, the truncation of the Hilbert space might
lead to deviations from the above asymptotic behavior. \cite{Wang}
The deviations can be substantial for small gaps, especially when 
the dimension of the Hilbert space used in the computation
is not sufficiently large.\cite{Schollwock} 

In our computation, the largest truncation errors are of the order of 
$10^{-7}$ for the conventional s-wave limit, while for the biggest 
$J_d$ values considered, the errors are as big as  $10^{-3}$. As the ratio 
$J_d/J_s$ increases, the truncation error becomes larger since the 
Kondo screening cloud extends across substantially longer distances and 
the magnetic fluctuations grow.  As a result, the accuracy of our 
calculation is significantly reduced in this limit.

\begin{figure}[h]
 \begin{minipage}{\linewidth}
 \centerline{\epsfxsize8cm\epsffile{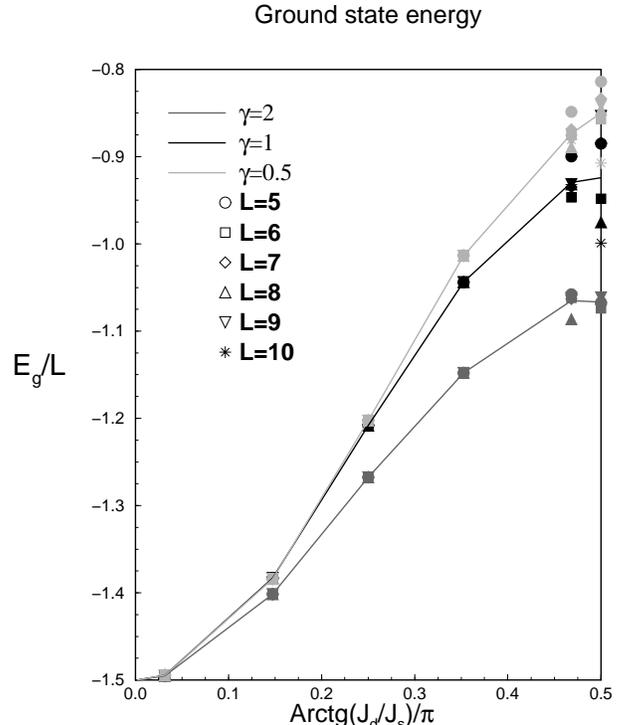}}
\caption[]{Ground state energy per rung of the two-channel Kondo 
ladder model versus $\arctan{(J_d/J_s)}/\pi$ for transverse
exchange parameter $\gamma=2$ (dark gray 
symbols), $\gamma=1$ (black 
symbols) and $\gamma=0.5$ (light gray 
symbols). We have plotted the results for ladders of 5 (circles), 
6 (squares), 7 (diamonds), 8 (up-triangles), 9 (down-triangles) and 
10 (stars) sizes. Note that the lines in this plot are only 
guides to the eyes.}
\label{fig:grenergy}
\end{minipage}
\end{figure}

\section{\bf Ground state energy and spin and charge excitation gaps}

Let us define the ground-state energy in a given spin-S subspace for 
a finite ladder with L-rungs by $E_g(L,N_c,S)$, where $N_c$ is the 
number of conduction electrons. Given that we are interested in the 
insulating ground state, we will consider a half-filled band with one 
electron per site and a total of $N_c= 2 L$ electrons.  The spin gap 
is defined as the energy difference between the ground state and 
the lowest spin excitation that changes the total spin quantum number 
by one:
$\Delta_s(L)=E_g(L,2L,S=1)- E_g(L,2L,S=0)$.

Similarly, the charge gap is the energy difference between the ground 
state and  the lowest pure charge excitation, which changes the total 
carrier number by two and keeps the spin quantum numbers fixed,
$\Delta_c(L)=E_g(L,2 L+2,0)- E_g(L,2 L,0)$.

Fig. \ref{fig:grenergy} displays the energy of the ground state as a 
function of $\arctan{(J_d/J_s)}/\pi$ for the three values of $\gamma$ 
explored in our calculation.  Results for several lattice sizes have 
been plotted. The ground state energy increases with the strength of 
the d-wave coupling for any value of $\gamma$ and  reaches the highest
values when   $\gamma=0.5$.

Let us now discuss the results for the ground state and excited state
energies of the conventional Kondo lattice model, $J_s=1$ and  $J_d=0$.
In the strong coupling limit, the energies can be calculated by 
using a perturbative expansion in $t/J$.\cite{Ueda} 
To second order in this small parameter, the ground state energy per 
rung is $E_g=-(3/2) J -(4 t^2/J)=-1.5004 $, the spin gap is 
$\Delta_s =J-(20 t^2/J)=0.998 $, and the charge gap is 
$\Delta_c=(3/2) J- 4 t +(9 t^2/J)=1.4609$, where we have used the 
parameters of our model.  Our numerical results are: 
$E_g=-1.50040$, $\Delta_s=0.9981$
and $\Delta_c=1.4614$, where we have used the scaling discussed in the 
previous section.  Given the almost complete agreement between the 
perturbation expansion and our numerical results, the accuracy of our 
computation in this limit is confirmed.

In the d-wave limit, $J_s=0, J_d=1$, the ground state energies  for the 
three different values of $\gamma$ are the following:
\bea
 \bullet \hspace{0.1in} \gamma &=& 2, 
E_{ground}^{rung} \sim -1.05  \nonumber \\
 \bullet \hspace{0.1in}\gamma &=& 1, E_{ground}^{rung} 
\sim -0.93 \nonumber \\
\bullet \hspace{0.1in}\gamma &=& 0.5, E_{ground}^{rung} 
\sim -0.88 \nonumber 
\eea
These are reasonable values for the ground state energy in the d-wave 
strong coupling limit. For large $\gamma$, the system will tend to 
form singlets along the rungs of the ladder. The energy of such 
configuration is: 
$E_{ground}^{rung} = 2 \frac {\gamma^2}{2+\gamma^2} (-3/4)$. 
For $\gamma=2$, the ground state energy is just $E_{ground}^{rung}=-1$.
The value obtained in our calculation is rather close to this simple
limit. In the opposite case of $\gamma=0$, let $t=0$, there are four
decoupled chains.  The spins at each chain tend to form spin-singlets
with their two neighboring electrons with exchange coupling $J_d/2=1/2$.
Therefore, the ground state energy per rung is related with the
ground state energy of four Heisenberg chains:
$E_{ground}^{rung} = 4 (1/2) (1/4 - \ln 2)=-0.886$.
The value obtained for $\gamma=0.5$ in our calculation is again
close to this simple limit.
The numerical ground state energy for $\gamma=1$ is
between the upper bound of $-1$ and the lower bound of
$-0.886$. Therefore, our result agrees with this simple 
physical argument.

\begin{figure}[h]
 \begin{minipage}{\linewidth}
 \centerline{\epsfxsize8cm\epsffile{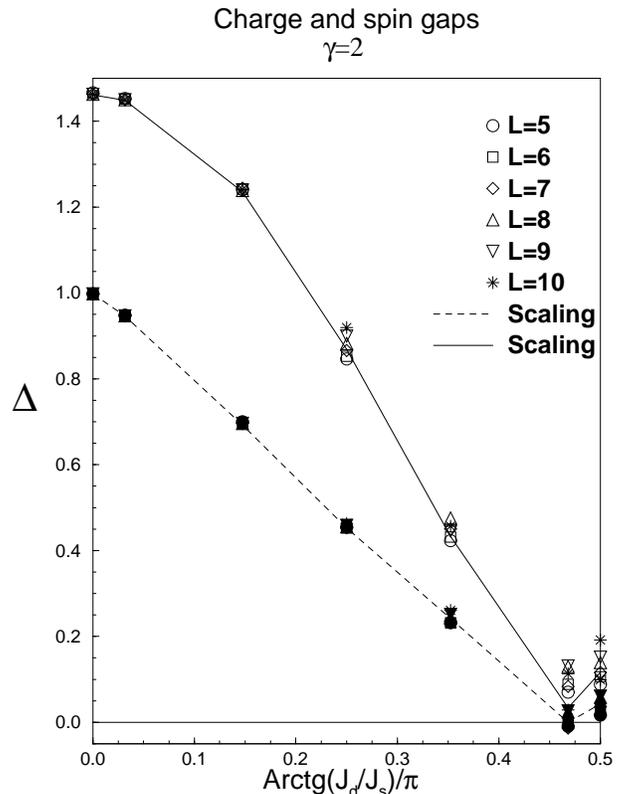}}
\caption[]{Charge gap (white symbols) and spin gap (black symbols)
versus $\arctan{(J_d/J_s)}/\pi$ for a d-wave 
transverse exchange parameter $\gamma=2$. We have plotted 
the results for ladders of 5 (circles), 6 (squares), 7 (diamonds),
8 (up-triangles), 9 (down-triangles) and 10 (stars) sizes. 
Also, the charge (solid line)  and spin (dashed line) gaps  obtained
using the scaling procedure are shown.}
\label{fig:gap2}
 \end{minipage}
\end{figure}

\begin{figure}[b]
 \begin{minipage}{\linewidth}
 \centerline{\epsfxsize8cm\epsffile{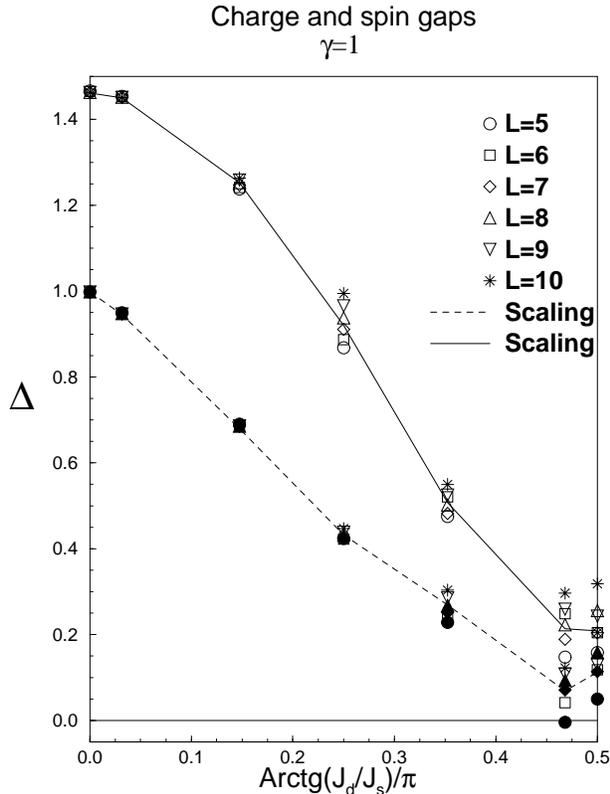}}
\caption[]{Charge gap (white symbols) and spin gap (black symbols)
versus $\arctan{(J_d/J_s)}/\pi$ for a 
transverse exchange parameter $\gamma=1$.
Notations are the same as before.} 
\label{fig:gap1}
 \end{minipage}
\end{figure}

\begin{figure}[h]
 \begin{minipage}{\linewidth}
 \centerline{\epsfxsize8cm\epsffile{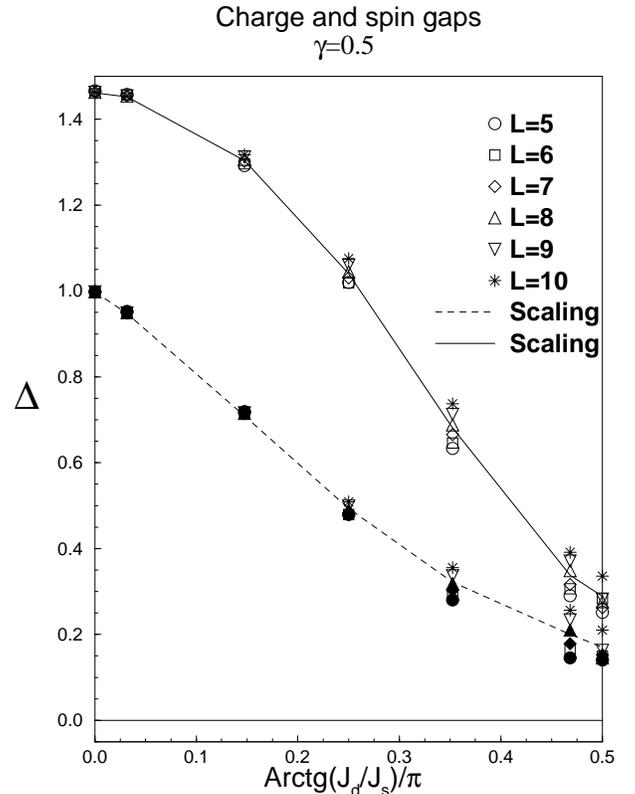}}
\caption[]{Charge gap and spin gap 
versus $\arctan{(J_d/J_s)}/\pi$ for a 
transverse exchange parameter $\gamma=0.5$.
Notations are the same as before.}  
\label{fig:gaph}
 \end{minipage}
\end{figure}

Figures \ref{fig:gap2}, \ref{fig:gap1} and \ref{fig:gaph} display 
the charge and spin gaps versus $\arctan{(J_d/J_s)}/\pi$ for  
$\gamma=2$, 1 and 0.5, respectively.
Results for several lattice sizes have been plotted. The gap 
resulting from using the scaling law (\ref{scaling}) is also shown. 
In addition, the first column of tables \ref{tab:spin} and
\ref{tab:charge} displays  the spin and charge
correlation lengths, respectively, derived from the scaling parameters of 
equation (\ref{scaling}), $\xi=v/\Delta(L=\infty)$. 

The behavior of the spin and charge gaps for any value of the 
parameter $\gamma$ is  similar. By increasing the d-wave coupling, 
the system evolves to a state with much smaller spin and charge gaps
for {\it any} $\gamma=2,1,0.5$.  
Comparing figures \ref{fig:gap2}, \ref{fig:gap1} and \ref{fig:gaph},
it can also be seen that $\Delta_s$ 
and $\Delta_c$ in the d-wave limit increase  as $\gamma$ decreases. 
So, a transverse exchange parameter of $\gamma=2$ is the most 
favorable value to develop a gapless region in the phase space. 
In fact, the smallest values of the gaps correspond to $\gamma=2$ and 
$J_d/J_s=10$, where by using scaling (\ref{scaling}) we are able to 
infer values of $\Delta_{spin}=2 \cdot 10^{-7}$ and 
$\Delta_{charge}=0.034$. Our belief is that an increase in  values of 
the d-wave Kondo exchange coupling, which implies a delocalization
of the singlet formed between the spin and the conduction electrons, 
induces a reduction of the energy gaps.
We think that for $\gamma <1$, the ladder sustains a spin and 
charge gap. However, for $\gamma >1$, the system develops gapless 
excitations  at a finite value of the ratio $J_d/J_s$.
Unfortunately, the accuracy of the calculation is not enough to get 
precise results in the $J_d/J_s \rightarrow \infty$ limit. We can not 
conclusively establish our expectation that a gapless region is 
present for any value of the ratio $J_d/J_s \gtrsim 10$ when
$\gamma >1$.
Also, we do not  know whether spin and charge gaps drop to zero 
precisely at the same ratio of $J_d/J_s$ or the spin excitation becomes 
gapless before the charge gap closes. 

Other models 
also display gapless excitations at  half-filling, such
as the attractive
Kondo-Hubbard model, where, in addition to the Kondo exchange, the 
conduction electrons have an on-site attraction.\cite{Kondo-Hubbard}
In the strong coupling limit of that model, there is a range of 
parameters  where both excitation gaps become equal and then drop to 
zero. However, in our model, although both excitation gaps are 
strongly reduced, the ratio between them grows when $J_d/J_s$ 
increases. 
This fact implies that 
the  time scale of the collective spin fluctuations that destroy the 
antiferromagnetic long-range order increases faster with $J_d/J_s$ 
than the characteristic time scale of the conduction-electron 
propagation. 

Let us conclude this Section commenting on the numerical difficulties 
of the problem.  The computation is much more involved in the d-wave limit 
($J_s=0$, $J_d=1$). 
The gap is, at least, two orders of magnitude smaller. 
Also, the correlation length in the ground state
increases roughly by two orders of magnitude, and longer ladders are 
required to reach the convergent regime with respect to the chain 
length. The number of degrees of freedom in the Kondo lattice model
is eight states per site. Therefore, the exact diagonalization can be 
carried out only for rather short ladders (five  rungs in our case).

In addition, as it is the case for  the one dimensional s-wave  Kondo 
lattice, \cite{Yu93} we  find that the Ruderman-Kittel-Kasuya-Yosida 
(RKKY) interaction increases in importance as $J_s$ decreases. For 
small $J_s$, large $J_d$, the system is so close to an instability 
that the numerical inaccuracy associated with larger ladder sizes is 
able to induce spurious antiferromagnetic ordering of the lattice, 
because the states with and without long range order are very close in 
energy. 

Also, in the d-wave limit, the smallness of the hopping term 
makes difficult  the redistribution of the local densities and, as
consequence, the determination of the smooth and 
energetically lowest electronic states.
Our approach to  improve the numerical accuracy has been to make 
several self-consistent sweepings for each block to guarantee the 
smooth redistribution of the local densities.

\section{\bf Spin and charge correlations in the ground state. 
Characteristics of the spin excited states}

\begin{figure}[h]
 \begin{minipage}{\linewidth}
 \centerline{\epsfxsize8cm\epsffile{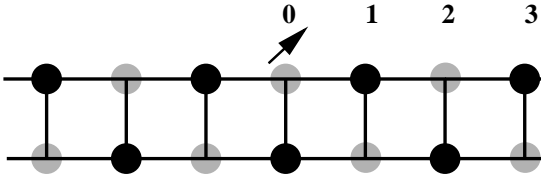}}
\caption[]{The black 
sites couple with the spin  through the d-wave Kondo 
channel. The gray 
sites couple through the conventional s-wave channel.}
\label{fig:zig-zag}
 \end{minipage}
\end{figure}

The DMRG method enables us 
to calculate the equal-time spin and charge correlation 
functions. 
We minimize the effects of the open boundary by 
evaluating
correlations between  the central site 
and  other sites along the  ladder. 
To clarify the nature of the correlations, we consider the pure d-wave 
exchange model with no hopping ($t=0$). In this case, the ladder 
decouples in two disconnected zig-zag chains. If only d-wave exchange 
is present, the spin sitting at position ``0'' in 
Fig. \ref{fig:zig-zag} correlates {\it only} with the electrons at the 
black 
sites, not with the ones in the gray 
chain.  Correlations 
between the central spin and electrons at the gray 
sites are induced 
by either s-wave exchange or hopping terms.  We use 
black and gray 
colors to display the different correlation functions. For example, 
the spin-spin correlation $\langle \vec S^0_0 \cdot \vec s_i^k \rangle$
between the central spin and the electron at site $\vec R_i^k$ will be 
display in black 
if $\vec R_i^k$ belongs to the black 
chain and in gray 
in the opposite case.

In the calculation of the correlation functions, we have used the 
local spin symmetry of the ground state 
($\langle \rho_{i\uparrow}\rangle=\langle \rho_{i\downarrow}\rangle$)
and the fact that the ladder is half-filled.\cite{approx}  We only 
discuss results for a d-wave transverse exchange value of 
$\gamma=2$, but the correlations for the other values of $\gamma$ 
exhibit similar characteristics.

\begin{figure}[h]
 \begin{minipage}{\linewidth}
 \centerline{\epsfxsize8cm\epsffile{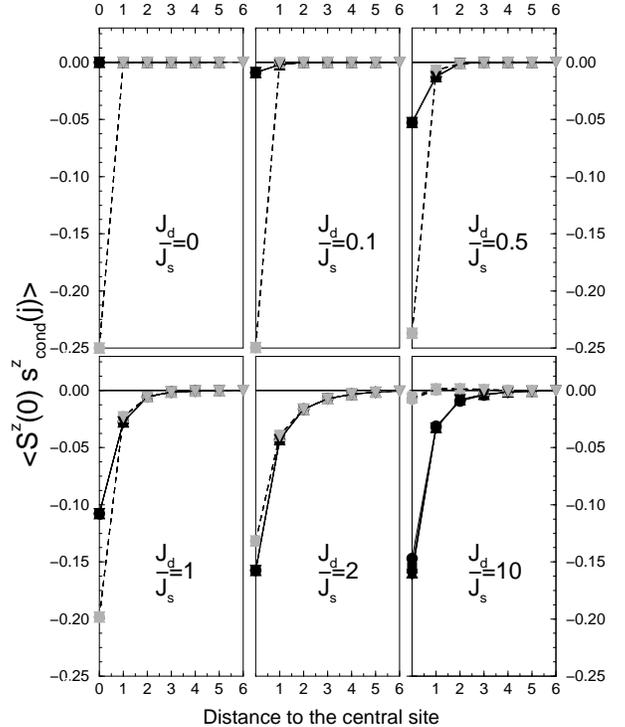}}
\caption[]{Spin-spin correlation between the central spin and 
the electron  at rung {\it j} as a function of the distance {\it j}. 
Correlations with electrons at gray (black) 
sites are shown in 
gray (black). 
The transverse exchange is $\gamma=2$.  Values 
for several ladder sizes are represented with different symbols 
as in previous figures.} 
\label{fig:ScS}
 \end{minipage}
\end{figure}

Figure \ref{fig:ScS} shows the spin-spin correlation between the local 
moment at the center of the ladder and  electrons at different sites, 
$\langle \vec S_0 \cdot \vec s_i \rangle$, as a function of the 
electron position for several values of the ratio $J_d/J_s$. It can be 
seen how the ground state evolves 
with increasing  values of  $J_d/J_s$ 
from a conventional insulating state with strong on-site 
correlations to a new state with large spin-spin correlation along 
the rungs and longer correlation lengths.

In the s-wave limit (top-left graph in Fig. \ref{fig:ScS}) the value 
of the on-site correlation is as large as
$\langle {S^{z}}^ 0_0   {s^{z}}^0_0 \rangle=-0.2499$.  
The correlations with the nearest neighbors in this case 
are extremely small,
$\langle  {S^z}^0_0  {s^z}^0_1 \rangle \sim
\langle  {S^z}^0_0  {s^z}^1_0 \rangle \leq 0.001$. 
This confirms the localized character of the ground state 
when only the  on-site exchange is present.
As $J_d/J_s$ grows,  the correlation with the electrons at black 
sites gradually increases, and the correlation with the gray 
sites decreases. When only d-wave exchange is present ($J_s=0$, $J_d=1$) 
the correlation with the electrons in the black 
chain  becomes dominant. 
The correlation with the electron along the rung reaches its highest 
value of $\langle  {S^z}^0_0  {s^z}^1_0 \rangle=-0.16$.

\begin{figure}[t]
\begin{minipage}{\linewidth}
\centerline{\epsfxsize8cm\epsffile{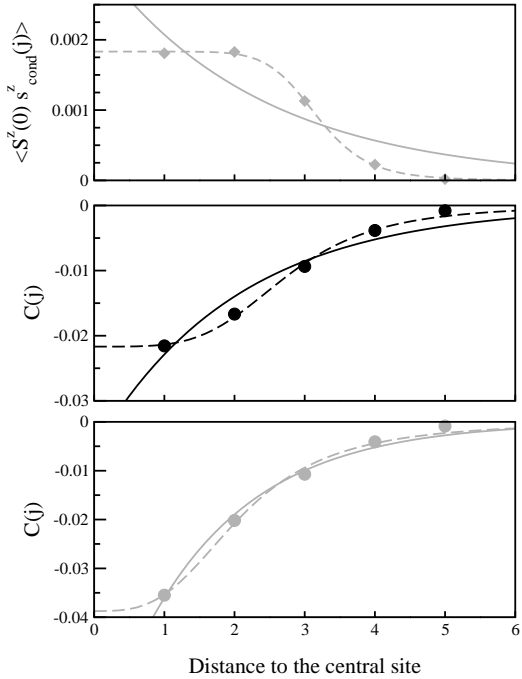}}
\caption[]{Comparison between exponential and power law
fits to the  DMRG results for the correlation as function
of the distance to the central site for $J_d/J_s=10$ and a ladder
of $10$ sites. Top graph, middle graph and bottom graph display the 
numerical data (circles), exponential (solid line) and power law 
(dashed line) fits for the spin correlation
along the gray chain,  the charge correlation along the black
chain and the charge correlation along the gray chain, respectively}
\label{fig:fit}
\end{minipage}
\end{figure}

We are also able to infer typical values of the spin correlation 
lengths by fitting the spin-spin correlation functions. 
An exponential  fit, 
$\langle \vec S_0 \cdot \vec s_j \rangle= C \exp{[-j/\xi^{spin}]}$, 
is appropriate for most of the ratios $J_d/J_s$. However,
for $J_d/J_s\gtrsim 10$ a power law fit,
$\langle \vec S_0 \cdot \vec s_j \rangle= C/[(j/\xi^{spin})^{\nu}+1]$,
more accurately describes the correlations along the gray 
chain, as it can be seen in the top graph of Figure (\ref{fig:fit}).
The transition from an exponential to a power law decay suggests the 
development of gapless spin excitations in the system,
in agreement with  our previous conclusions. 
Table \ref{tab:spin} displays the correlation lengths between the 
central moment and the electrons at the gray (black) 
zig-zag chain, 
$\xi^{spin}_{gray
}$ ($\xi^{spin}_{black
}$), for
different values of the ratio $J_d/J_s$.
The  correlation length derived from the scaling of the excitation
gap  ($\xi^{spin}=v/\Delta$) is also shown.  
We notice that $\xi^{spin}$ and $\xi^{spin}_{gray
}$ have similar
values and both increase with the ratio $J_d/J_s$. 
On the other hand, $\xi^{spin}_{black
}$ is roughly constant for all the 
values studied. 

\begin{minipage}{\linewidth}
\begin{table}
\caption{Spin-spin correlation lengths for different values of the 
ratio $J_d/J_s$ (first column). The second column 
displays the lengths derived from
the scaling law. The third and fourth columns show 
the correlation lengths between the central local moment 
and electrons at the gray and black 
chains, respectively. }
\label{tab:spin}
\begin{tabular} {||c|c|c|c||} 
 $J_d/J_s$ & $\xi^{spin}=v/\Delta$    
                 & $\xi^{spin}_{gray
}$ & $\xi^{spin}_{black
}$ \\
\hline
0   &	0.02	 &	0.061    & 	0.67\\
0.1 &	0.13	 &	 0.14  	 &	0.66\\
0.5 &	0.33     &	 0.281   &	0.66\\
1   &	0.48	 &	 0.47	 &	0.72\\
2   &	0.71	 &	 0.88    &	0.81\\
10  &	 -       &       3.17	 &	0.64\\
\end{tabular}
\end{table}
\end{minipage}

\begin{figure}[t]
\begin{minipage}{\linewidth}
\centerline{\epsfxsize8cm\epsffile{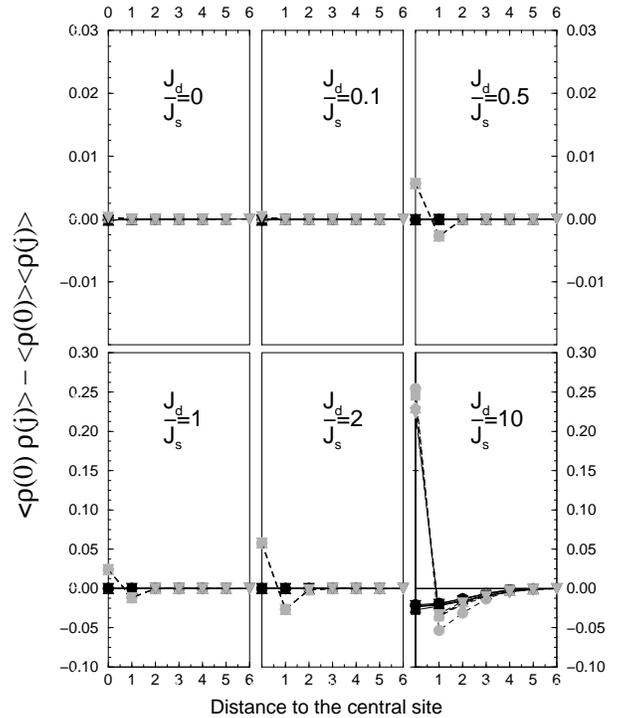}}
\caption[]{Dependence of the charge correlation function
on the  distance to the 
central site for a transverse exchange $\gamma=2$.  Correlations 
with electrons at gray (black) 
sites are shown in gray (black). 
Note the factor of ten difference in the vertical
scale between the top and 
bottom graphs. Values for several ladder sizes are presented with 
different symbols as in previous figures.} 
\label{fig:rr}
\end{minipage}
\end{figure}

Let us now discuss the characteristics of the charge
correlation functions.  Figure \ref{fig:rr} displays the 
charge correlation, 
$C(j)=\langle \rho(0) \rho(j)\rangle
- \langle\rho(0)\rangle \langle\rho(j)\rangle$,
as function of $j$, the position of the second electron.
The charge-charge correlation is practically zero everywhere for 
a pure s-wave exchange.  This behavior agrees with  the strong 
coupling limit of the Kondo model. The introduction of a finite 
d-wave coupling induces an on-site charge correlation that becomes 
quite big in the d-wave limit, 
$C(0)\sim 0.24$. The correlation 
with the remaining sites in the ladder is smaller and negative. This 
large on-site charge-charge correlation is related to the non-local
character of the singlet state in the d-wave channel, as we discuss
in the next Section.

We can also notice in Fig. \ref{fig:rr} that the correlation 
length for both zig-zag chains grows with $J_d/J_s$.
Table \ref{tab:charge} shows the correlation lengths between the 
electron at the central site and electrons in the gray (black) 
zig-zag chain, 
$\xi^{charge}_{gray
}$ ($\xi^{charge}_{black
}$), for
different values of the ratio $J_d/J_s$. Also, the  correlation 
length derived from the scaling ($\xi^{charge}=v/\Delta$) is shown.  
All the correlation lengths grow with the ratio $J_d/J_s$, but their 
precise values are different.  As in the spin case, the charge 
correlation lengths have been obtained using an exponential fit 
in the region $J_d/J_s < 10$ and a power fit for $J_d/J_s \geq 10$. 
Middle and bottom graphs in Figure (\ref{fig:fit}) show a comparison 
between exponential and power law fits for $J_d/J_s=10$. 

From the analysis of the correlation functions, we conclude that 
there is a phase transition in the two-channel Kondo lattice when
the transverse exchange parameter is larger than unity.
Spin and charge correlation 
functions evolve from the ones typical of 
a conventional Kondo lattice to those of a d-wave Kondo ladder with 
gapless spin and charge modes.
The value of the spin correlation length derived from the scaling
law (eq. (\ref{scaling})) correlates {\it only} with the spin
correlation length along the gray 
chain. The spin correlation length along 
the black 
chain appears to be determined by the hopping interaction rather than 
by the d-wave exchange coupling.
On the other hand, the charge correlation lengths of the two zig-zag 
chains (black and gray) 
behave similarly and are in agreement with the 
lengths associated with  the charge gap scaling.
In any case, we have to keep in mind that results for larger 
ratios of $J_d/J_s$ are likely to be less reliable 
given that for these values 
$2 \xi \sim L_{diag}$, where $L_{diag}=5$ is the length for which an 
exact diagonalization is carried out.

\vspace{0.2in}
\begin{minipage}{\linewidth}
\begin{table}
\caption[]{Charge-charge correlation lengths for different values of 
the ratio $J_d/J_s$. The second column displays the lengths derived from
the scaling law. The  third and fourth columns show the correlation lengths 
between the central site and sites in the gray and black 
chains, respectively.}
\label{tab:charge}
\begin{tabular}{||c|c|c|c||}
 $J_d/J_s$ & $\xi^{charge} =v/\Delta$ & $\xi^{charge}_{gray
}$ & 
  $\xi^{charge}_{black
}$ \\
\hline
0 &	0.146 	 &	0.43 &	0.67\\
0.1 &	0.150	 &	0.10 &	0.70\\	 
0.5 &	0.25	 &	0.17 &	0.95\\	 
1 &	0.34	 &	0.24 &	1.77\\ 
2 &	0.81	 &	0.36 &	1.32\\ 
10 &	8.56	 &	2.09 &	2.75\\
\end{tabular}
\end{table}
\end{minipage}
\vspace{0.2in}

\begin{figure}
 \begin{minipage}{\linewidth}
 \centerline{\epsfxsize8cm\epsffile{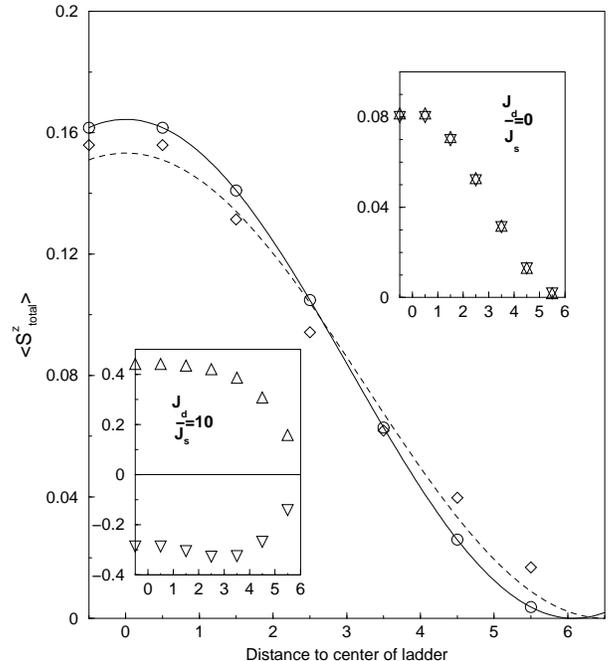}}
\caption[]{Third component of the total spin of  each rung as a 
function of the distance to the center of the ladder for $J_d/J_s=0$, 
open circles, and $J_d/J_s=10$, open diamonds.
The lines represent the best fit to the expression:
$S^z_{total}=a_0 (\sin{[\pi (j+(L_{open}/2))/L_{open}]})^2$.
The solid line fits the $J_d/J_s=0$ points, dashed line fits 
the $J_d/J_s=10$ set of points. The insets display the 
z-component of the spin of the local moments (triangles up)
and the conduction electrons (triangles down) for these 
two values of the ratio between the exchange couplings.}
\label{fig:spinstate}
 \end{minipage}
\end{figure}

Let us conclude by discussing some characteristics of the spin excited 
states. Fig. \ref{fig:spinstate} displays the value of the third 
component of the total spin in each rung as a function of the distance 
between the rung and the center of the ladder for $J_d/J_s=0$, open 
circles, and $J_d/J_s=10$, open diamonds. In the insets, the 
z-component of the spin of the local moments, $\langle S^z_{j}\rangle$,
and the conduction electrons, $\langle s^z_{j}\rangle$,
are displayed.  The results correspond to a ladder of 12 sites with 
open boundary conditions.

In the case of the conventional s-wave limit, the spin of the local 
moments, $\langle S^z_{j}\rangle$,  and of the electrons, 
$\langle s^z_{j}\rangle$, are almost identical for any site 
in the ladder, as it can be seen in the top inset of 
Fig. \ref{fig:spinstate}. The magnitude of the spin is also equally 
distributed between the lower and upper chains.
The local and electronic spins follow a square sine function,
appropriate for a massive mode,\cite{Sorensen}
\bea
  \langle S^z_{j,total}\rangle = \langle S^z_{j}+ s^z_{j}\rangle
\sim 2 \langle S^z_{j}\rangle \sim 
			2 \langle  s^z_{j}\rangle \sim  \nonumber \\
\frac{1}{L_{open}} 
	\big[ \sin{\frac{\pi (j+(L_{open}/2))}{L_{open}}} \big]^2 , 
\eea
where $j$ is the position under consideration and $L_{open}$ is the 
length over which the spin excitation extends, which is longer than 
the ladder length. By making a two-parameter fit, 
$\langle S^z_{j,total}\rangle = 
C \big[ \sin{\frac{\pi (j+(L_{open}/2))}{L_{open}}} \big]^2$ we infer 
$L_{open}=12.16$ and a normalization coefficient  
C=0.0822, in  agreement with the value of $1/L_{open}$.
This agreement can also be clearly seen in Fig. \ref{fig:spinstate}, 
where the solid line is the fitting function.

In contrast with the case of the conventional Kondo ladder, the local 
and the conduction spins display very different  behavior in the 
d-wave limit (see the bottom inset in Fig. \ref{fig:spinstate}).  The third 
component of the local spin of each rung
is always positive; it decreases from 
the center ($\langle S^z_0 \rangle\sim 0.44$) to the end site 
of the ladder ($\langle S^z_L \rangle \sim 0.16$).  On the other hand, 
the magnitude of the z-component of the electronic spin is always 
negative and ranges between  
$-0.28$ and $-0.14$.
However, the sum of both spin densities,  
$\langle S^z_{j,total}\rangle$ behaves similarly to the case of only 
on-site exchange coupling, although it is clear that the dispersion 
of the data is larger (Fig. \ref{fig:spinstate}).
By using the same two-parameter fitting,
$\langle S^z_{j,total}\rangle = 
C \big[ \sin{\frac{\pi (j+(L_{open}/2))}{L_{open}}} \big]^2$, 
we obtain a length of $L_{open}=13.02$ and a normalization coefficient 
C=0.0766, which should be compared with $1/L_{open}=0.0768$.

\section{\bf Variational calculation}

In order to gain more insight into the nature of the 
ground state in the d-wave limit, we  have performed a variational 
calculation.  Variational wave functions have been successfully used 
in the conventional Kondo lattice model.\cite{Yosida}.
We restrict ourselves to a simpler Hamiltonian with only  d-wave 
coupling. In this case, as we have discussed previously, the ladder 
decoupled in two disconnected chains (Fig. \ref{fig:zig-zag}). 
We can label the sites of one of the zig-zag chains just by their $j$ 
coordinate. So, the Hamiltonian for one of the decoupled chains is 
reduced to:
\beq
H_d=\frac{1}{2(2+\gamma^2)}\sum_{j} \vec S_j \cdot
({\bf d}^{\dg}_{j, s}(-\gamma)\vec \sigma_{s s^{'} } 
{\bf d}_{j, s^{'}}(-\gamma))
\label{eq:d-wave-Kondo}
\eeq
with
${\bf d}^{\dg}_{j, s}(x)=\psi^{\dg}_{j+1,s}+\psi^{\dg}_{j-1,s}
+x \psi^{\dg}_{j,s}$, 
where $\psi^{\dg}_{j,s}$ creates a conduction electron with 
spin $s$ on site $j$.

Our variational wave function is the product of a Zhang-Rice-like
singlet at each site: 
\beq
\vert \Psi_0\rangle= \prod_l \big(
\vert \uparrow \rangle_{l} {\bf d}^{\dagger}_{l,\downarrow}(\alpha)
- \vert \downarrow \rangle_{l} {\bf d}^{\dagger}_{l,\uparrow}
(\alpha)\big)
\vert 0\rangle,\\
\label{eq:varwave} 
\eeq
where $\vert \uparrow \rangle_{l}$ ($\vert \downarrow \rangle_{l}$)
indicates that the lth spin is in the state with 
$S_z=+1/2$ ($S_z=-1/2$). 
The operator ${\bf d}^{\dagger}_{l,\sigma}(\alpha)$
creates a non-orthogonal Wannier state
which overlaps with the states at neighboring sites. We allow
$\alpha$ to vary over values that differ from $\gamma$. 

By decomposing $\alpha$ into its magnitude and its phase, 
$\alpha=|\alpha|e^{i\phi}$, we find that the energy minimum 
for any transverse exchange parameter ($\gamma$) and 
length of the chain is always reached when $\phi=\pi$. 
Table \ref{tab:variational} displays the 
values of $|\alpha|$ for which the variational energy becomes minimum and 
the value of that minimum.
For $\gamma=2$, the asymptotic minimum  is
in reasonable agreement with the computed value of $-1.05$.
The variational energy obtained for $\gamma=0.5$ is also 
remarkably close to the value obtained in the DMRG calculation 
($E_{ground}^{rung} \sim -0.88$).
However, the variational energy obtained for $\gamma=1$ is too high;
it is even higher than the energy of the configuration when
half of the spins form singlets with one of their nearest neighbors
($E_{ground}^{rung}=-3/4$). 
The energy minimum  for $\gamma=2$ is a very shallow one, and there
is a large region of $|\alpha|$ values where the energy is very close 
to $-1$. In contrast, for $\gamma=1$ and $0.5$ the energy minimum is
reached at finite values of the parameter $|\alpha|$, and a scaling 
in $1/L$ is necessary to obtain the ground state energy of 
an infinite system.

\vspace{0.2in}
\begin{minipage}{\linewidth}
\begin{table}
\caption[]{Results of our variational calculation for the 
studied transverse exchange parameters. 
The second and third columns show the magnitude of $|\alpha|$ where 
the energy minimum is achieved and its value, respectively.}
\label{tab:variational}
\centerline{\begin{minipage}{0.7\linewidth}
\begin{tabular}{||c|c|c||}
 $\gamma$ & $|\alpha|$ & $E_{ground}^{rung}$  \\ 
\hline
2&	$\infty$ &  -1.\\
1 &	  0.45 &	-0.71\\	 
0.5 &	0.26	 & -0.89\\	 
\end{tabular}
\end{minipage}}
\end{table}
\end{minipage}
\vspace{0.2in}

\begin{figure}[h]
 \begin{minipage}{\linewidth}
 \centerline{\epsfxsize8cm\epsffile{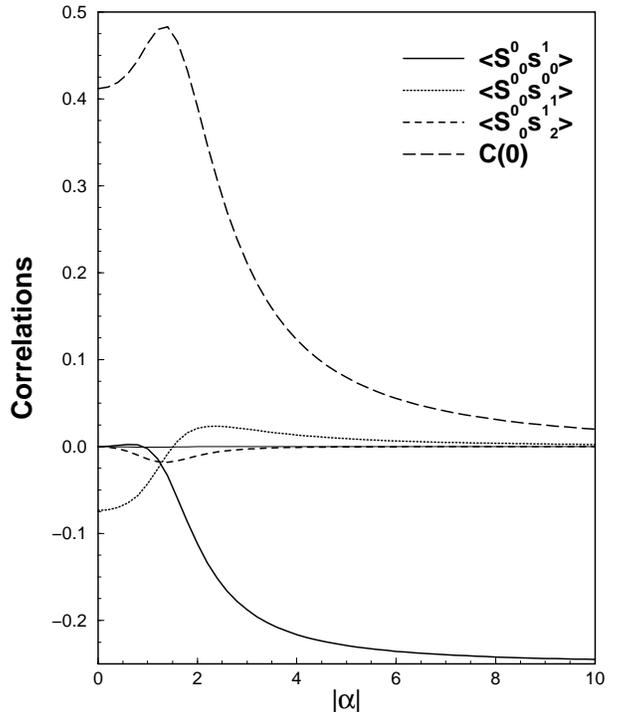}}
\caption[]{Correlations of variational wave function (\ref{eq:varwave})
versus the parameter $|\alpha|$. Spin correlation between the central 
moment and the electron along the rung 
($\langle  {S^z}^0_0  {s^z}^1_0 \rangle$), solid line; correlation
with the nearest electron along the chain  
($\langle {S^z}^0_0  {s^z}^0_1 \rangle$), dotted line, and
correlation with the next-next-nearest electron along the rung
($\langle  {S^z}^0_0 {s^z}^1_2 \rangle$),  dashed line.  The 
on-site charge-charge correlation, 
$C(0)=\langle \rho(0) \rho(0)\rangle- 
\langle \rho(0)\rangle \langle\rho(0)\rangle$, 
is also displayed (long-dashed line).}
\label{fig:varcorr}
 \end{minipage}
\end{figure}

We have also calculated the charge-charge and spin-spin correlations
as functions of $|\alpha|$, for  $\phi=\pi$. The spin-spin correlation
between the local spin at the center of the ladder and the 
two nearest, and next-next-nearest  electrons sitting in the 
black 
chain are displayed in Fig. \ref{fig:varcorr}, together with the 
on-site charge-charge correlation at the central spot.
Since gray and black 
chains are totally decoupled in our model 
(Hamiltonian (\ref{eq:d-wave-Kondo})), correlations with electrons at 
the gray 
sites are absent. From this plot we conclude 
that extended Kondo models, where the moments
couple with electrons on  {\it other}  (rather than it's own) sites,
display extended spin correlations  and strong on-site charge-charge
correlations. There is a large range of values of the variational 
parameter $|\alpha|$ where the spin and charge correlations 
reach big values. For example, for $|\alpha|= 2.8$, 
$C(0)=0.24$  and  
$\langle  {S^z}^0_0  {s^z}^1_0 \rangle=-0.18$. These values 
are in good agreement with those found in our DMRG calculation for 
$\gamma=2$. So, it is reasonable to think that the ground state 
of the d-wave Kondo ladder, at least for some values of the 
transverse coupling, can be understood in terms of extended 
singlets on each lattice site 
(Eq. \ref{eq:varwave}).

\section{\bf Conclusions}

Motivated by experimental
\cite{DMET2FeBr4}
and theoretical considerations about the phase diagram of 
Kondo lattice models, \cite{Piers} we have  used the density matrix 
renormalization group method \cite{White}  to study the strong 
coupling limit of a Kondo lattice model with two different 
interactions: the conventional  on-site Kondo exchange ($J_s$) and a 
d-wave exchange coupling ($J_d$). Our purpose has been to map  the 
phase diagram of this model and sort out whether or not  it displays 
a phase transition. 

By  increasing the ratio between the two couplings $J_d/J_s$, the 
system smoothly evolves from a conventional Kondo insulator
ground state, where a local singlet is formed at each site,  
to a ground state with much smaller spin and charge excitation 
gaps. For small values of $\gamma$ less than unity, our results
suggest that the system preserves a spin and charge gap for all
values of the $J_d/J_s$ ratio. However, 
when the parameter $\gamma$ is increased to values larger
than unity, there is clear evidence to 
suggest that the system develops gapless excitations 
for $J_d/J_s \sim 10$, suggesting a phase transition. 
For a transverse exchange parameter of $\gamma=2$ and 
within the accuracy of the calculations, both the
spin and charge gap become zero at this point. 
However, numerical difficulties 
related to the proximity to an antiferromagnetic instability preclude 
us from resolving definitively whether these gaps
are zero for $J_d> 10J_s$ corresponding to a phase transition into
a metallic phase, or whether the gaps re-establish
small but finite values at $J_d>10J_s$, corresponding to a critical
point at $J_d/J_s=10$. 
Our intention is to increase the Hilbert space and the length of the 
ladder to be able to get more accurate values of the excitations gaps 
in this limit.

This semimetallic ground state  
with gapless spin and charge excitations 
can be understood as a superposition of the extended  
singlets formed between each local moment and its three nearest-neighbor 
electrons. 
The characteristics of the ground state excitations change with
$J_d/J_s$. In particular, we find that  as $J_d/J_s$ rises, 
the ratio of charge to spin gap $\Delta_c/\Delta_s$ is enhanced. 
Simultaneously, the system develops strong 
spin-spin correlations between the local moment and the electron
along the rung and 
large on-site charge-charge correlations. The typical 
correlation lengths increase with the $J_d/J_s$ ratio. 
Also the transition from an exponential to a power law decay 
in the spin and charge correlations for $\gamma=2$ and 
$J_d/J_s > 10$ supports the existence of a gapless phase.

Additional insight into the  results at large $J_d$ has been obtained
from a variational calculation. 
By writing down a variational wavefunction in which 
an extended singlet forms 
between each moment and the three neighboring electrons in the ladder,
we obtain variational ground state energies that are 
close to our DMRG values for $\gamma=2$ and $0.5$. 
Our variational ansatz also displays large spin-spin correlations 
along the rungs and strong on-site charge-charge correlations as 
found in the DMRG results.

It would be interesting in future work to examine the effect of
three-leg and four-leg ladders on the above results.  By increasing
the number of legs, the full point-group symmetry of the
two-dimensional limit is gradually restored, permiting d-wave singlets
to delocalize without admixing into extended s-wave pairs. This may
permit the gapless phase to be reached at lower values of the ratio
$J_d/J_s$. Indeed, the vanishing spin and charge gaps observed in our
ladder model may become a full-fledged phase transition in truly
two-dimensional systems.  \cite{Piers}

\vspace{0.2in}
We are grateful for the support provided by the Abdus Salam 
International Center for Theoretical Physics (Trieste, Italy),
where this work was initiated and all numerical calculations 
were carried out. 
We thank Natan Andrei, Daniel Cabra, Andreas Honecker, Andres Jerez, 
Karine Le Hur, Edmond Orignac and Alexei Tsvelik 
for valuable discussions. 
We thank John Cooper for mentioning to us the earlier work of 
J.B. Dunlop (Ref.\onlinecite{Zn13X,Al11Mn4}) and Victor Yakovenko for 
pointing out Ref. \onlinecite{DMET2FeBr4}. 
This research was  partially supported by NSF grants DMR 9705473, 
DMR 9972087 and DMR 91-20000 through the Science and Technology Center 
for Superconductivity.  S. Qin was partially supported by the 
Chinese NSF. P. Coleman is supported by NSF grant DMR 9983150.

\end{multicols}

\begin{references}
\bibitem{char-DMET} While in the low temperature insulating phase the 
	magnetic properties are described in terms of the $Fe^{3+}$ 
	localized spins, in the high temperature metallic phase the 
	Curie-Weiss magnetism is described in terms of both $DMET^+$ 
	spins and $Fe^{3+}$ spins.  This behavior is in contradiction 
	to the case of ordinary metals where the conduction electrons 
	give rise to a small Pauli paramagnetic susceptibility without
	localized magnetic moments.
\bibitem{approx}We calculate the spin-spin correlation function to be:
\bea
\langle \vec S \cdot \vec s_i\rangle=
3\langle  S^z s_i^z\rangle=
3  \langle  S^z\frac {(\rho_{i\uparrow}-
\rho_{i\downarrow})}{2} \rangle= \nonumber\\
=3  \langle  S^z (\rho_{i\uparrow}-\frac{1}{2})\rangle
=3 \langle  S^z \rho_{i\uparrow}\rangle.\nonumber
\eea
The charge-charge correlations function is:
\bea
C(i)=\langle\rho_0\rho_i\rangle - \langle\rho_0\rangle \langle\rho_i\rangle
=\langle \rho_0 \rho_i \rangle - 1= \nonumber \\
=2[\langle(\rho_{0\uparrow}-1/2)(\rho_{i\uparrow}-1/2)\rangle+
  \langle (\rho_{0\uparrow}-1/2)(\rho_{i\downarrow}-1/2)\rangle].
	\nonumber
\eea
\end{references}
\end{document}